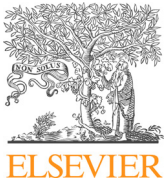
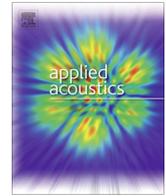

Technical note

# Reliability and repeatability of ISO 3382-3 metrics based on repeated acoustic measurements in open-plan offices

Manuj Yadav *, Densil Cabrera, James Love, Jungsoo Kim, Jonothan Holmes, Hugo Caldwell, Richard de Dear

*Sydney School of Architecture, Design and Planning, The University of Sydney, Sydney, NSW 2006, Australia*

ABSTRACT

This paper investigates variability in the key ISO 3382-3:2012 metrics, based primarily on the repeatability and reliability of these metrics, using repeated measurements in open-plan offices. Two types of repeated measurements were performed in offices – Type1 ($n$ = 36), where the same path over workstations was measured from opposite ends, and Type2 ($n$ = 7), where two different measurement paths were measured. Analyses performed per metric used (i) the *range* of observed values, i.e., $\Delta$Type1, $\Delta$Type2; and (ii) the *observed* values on their actual scales. Results from category (i) analysis: $\overline{\Delta\text{Type1}}$, and bootstrapped 95% confidence intervals were 1.2 m (0.9,1.5) for distraction distance ($r_D$); 0.8 dB (0.6,1.0) for spatial decay rate of speech ($D_{2,S}$); 1.2 dB (0.8,1.5) for A-weighted sound pressure level of speech at 4 m ($L_{p,A,S,4\ m}$); and 1.2 dB (0.7,1.7) for the A-weighted background noise level ($L_{p,A,B}$). $\overline{\Delta\text{Type2}}$ were between twice and thrice the respective values of $\overline{\Delta\text{Type1}}$. Results from category (ii) analysis: the *reliability*, based on intra-measurement correlation coefficients for repeated measurements, was fairly high for all metrics except for $L_{p,A,S,4\ m}$ for Type2 repeats. The *repeatability limit/coefficient* ($r$), which is the absolute difference between metric values not expected to be exceeded in 95% of the repeatability conditions, was 2.5 m for $r_D$; 1.7 dB for $D_{2,S}$; 2.9 dB for $L_{p,A,S,4\ m}$; and 3.0 dB for $L_{p,A,B}$, for Type1 repeats. The $r$ values for Type2 repeats were substantially higher except for $D_{2,S}$; $L_{p,A,B}$ not applicable in the current context. Overall, most of the Type1 results seem reasonable considering repeats were conducted in complicated room acoustic environments, while Type2 repeats would benefit from larger sample sizes in future studies. Some recommendations are outlined for the ISO 3382-3 methodology vis-à-vis Type1 and Type2 repeats, including future research directions that go beyond increased sample sizes.

Crown Copyright © 2019 Published by Elsevier Ltd. All rights reserved.

## 1. Introduction

First introduced in 2012, and unrevised as of yet, ISO 3382-3 [1] is the main international standard for measuring room acoustic metrics for open-plan offices. These metrics include those that are based on speech sound pressure level ($SPL_{Speech}$) – the spatial decay rate of speech per distance doubling ($D_{2,S}$), A-weighted SPL of speech at a distance of 4 m ($L_{p,A,S,4,m}$); distraction distance ($r_D$) based on spatial decay of the speech transmission index (STI)[1]; along with the A-weighted background noise level ($L_{p,A,B}$). Apart from $L_{p,A,B}$, the salient feature of these metrics is that they characterize aspects of the *spatial decay of speech over workstations* within an office[2]. These metrics are calculated from room acoustic measurements over a more-or-less linear *path*[3] comprising a set of measurement positions near workstations. As such, along with relevant room acoustic factors, ISO 3382-3 metrics also incorporate the effects of relevant design factors, including the furniture arrangement, the room geometry, etc. ISO 3382-3 requires at least two measurement paths per office. If only one measurement path is possible, then two measurements per path are to be conducted with the source loudspeaker positioned on the opposite ends of the path [1]. This study considers the variability in the key ISO 3382-3 metrics – $L_{p,A,B}$, $D_{2,S}$, $L_{p,A,S,4,m}$ and $r_D$ – due to repeated measurements in the same office (i) with two measurements that traverse the same set of workstations, but using the source loudspeaker on the opposite ends of the path per measurement, and (ii) with different measurement paths (see Section 2.3 for details).

---

\* Corresponding author at: Room 589, 148, City Road, Sydney School of Architecture, Design and Planning, The University of Sydney, NSW 2006, Australia.
*E-mail address:* manuj.yadav@sydney.edu.au (M. Yadav).
[1] Privacy distance and STI at the nearest workstation are also mentioned in ISO 3382-3, but are optional, and not considered here.

[2] For the purposes of this paper, and consistent with ISO 3382-3, an office is defined as an *acoustic zone* that is more-or-less uniform in its furniture arrangement, ceiling type, and horizontal sound absorption characteristics throughout.
[3] Use of the term 'line' is possible too, but 'path' is preferred here; also see section 2.3.

Haapakangas et al. [2] represents the only previous study where results regarding variability in the ISO 3382-3 metrics are provided. The results in Haapakangas et al. [2] were based on unpublished data from Hongisto et al. [3], wherein repeated measurements were conducted in a single open-plan office. In comparison, the current study considers variability in the ISO 3382-3 metrics using two types of repeated measurements, which were conducted within several open-plan offices (details in Sections 2.1–2.4, and Table 1).

## 2. Methods

### 2.1. Room acoustic measurements

Acoustic measurements were made according to ISO 3382-3 [1] in furnished, unoccupied offices with fully operational heating, ventilation, and air conditioning. These offices are either non-contiguous units within a larger floor plate, or single units that cover the entire floor. Overall, the measured offices included a reasonably wide variation in several relevant aspects, such as the floor plate geometries and areas, the partitions between workstations (ranging from the more traditional cubicle designs, to those with no partitions at all), ceiling heights and absorption treatment, density of workstations on the floor, etc. (Table 1). While achieving a more uniform spread of these aspects for statistical reasons would be desirable, the selection of offices in the current study was based mostly on pragmatic reasons – wherever access outside working hours could be negotiated with willing building managers. Nevertheless, strict quality control was enforced to select offices that are suitable for conducting ISO 3382-3 compliant measurements.

The software used to generate, play, record, and analyze (in post-processing) the signals was AARAE, which is a MATLAB-hosted graphical user interface driven environment [4]. The measurement signal used was an exponentially swept sinusoid (50 Hz – 20 kHz) of 30 s duration. The signal chain for the measurements included a computer running AARAE, an audio interface (RME Fireface UFX, Haimhausen, Germany), an amplifier (Brüel & Kjær Power Amplifier Type 2716-C, Nærum, Denmark), an omnidirectional loudspeaker (Brüel & Kjær OmniSource Type 4295) with a known sound power output that was calculated using ISO 3745 [5], omnidirectional microphones (Earthworks M30, Milford, USA), and a preamplifier unit (RME OctaMic XTC). For each path, the measurement signal from the source loudspeaker was recorded using between 5 and 8 microphones simultaneously, where the microphones were located at receiver positions near workstations, as described in ISO 3382-3 [1]. The number of microphones used within a given office was always the same – determined by the floor plate of the office measured – whereas the number of microphones used between offices varied. Background noise measurements were made at each microphone position for at least 60 s, except in some cases, where the extreme microphone position was not used for recording the background noise (see details in Section 2.3).

### 2.2. Calculation of ISO 3382-3 metrics

The sinusoidal sweep recordings from microphones per measurement path were used to derive the SPL$_{Speech}$ metrics, i.e., $D_{2,S}$ and $L_{p,A,S,4,m}$. Modulation transfer functions (MTFs) used for calculating STI, and subsequently $r_D$, used the male[4] octave-band weights and redundancy factors [8]. $L_{p,A,B}$ was calculated as described in ISO 3382-3, i.e., from octave-band SPLs that represented spatially averaged values for the microphones along a measurement path [1]. In order to avoid potential problems with extrapolation, $r_D$ values were considered acceptable if they were within the range of the measured physical distances extended by 10% on either end [8]. Otherwise, in order to use the remaining offices, the background noise was adjusted, with the same gain applied in all bands to all measurement paths in a given office, as described in [8].

While not a ISO 3382-3 metric, in order to further characterize the offices, spatiallyaveraged and mid-frequency (average of 500 Hz and 1 kHz octave-band values) reverberation times ($T_{30}$) were calculated from the impulse responses, using 12th-order filters [9].

### 2.3. Repeated measurements in offices

Variability in the ISO 3382-3 metrics was considered for two types of repeated measurements:

(a) **Type1**, where two measurements per path were conducted. Here, after conducting one measurement with the loudspeaker on one end of the path, the other measurement was conducted either by placing the loudspeaker on the other end as seen in Fig. 1, or by swapping the positions of the last microphone and the loudspeaker from the previous measurement. All the paths in the offices in Table 1 were measured using Type1 repeated measurements (36 in total). Note that most offices were measured using only a single path, and in these cases, the usage of 'path' and 'office' is interchangeable. However, the use of the term 'path' is preferred here over 'office' for offices with only one path, to avoid confusion.
(b) **Type2**, where two measurement paths were measured within the same office (i.e., the same acoustic zone). Type2 repeated measurements were only possible in a small subset of all offices measured (7 in total)[5]: these being the offices for which the allocated access times permitted several repeated measurements. The idea here was to determine the variation in the ISO 3382-3 metrics [1] over two paths that can be considered as likely options to characterize the office overall. These measurement paths included those that were on an adjacent row of workstations relative to the first path, for instance any two adjacent paths in Fig. 1, or more-or-less linear paths that were at a small angle to the length of the office, or the main measurement axis.

Note that due to time limitations, background noise measurements were not performed for all the repeated measurements per path. However, background noise was recorded for at least one of the Type1 repeated measurements, and for at least one of each path of the Type2 repeated measurements. Hence, calculation of $r_D$ was always possible, since it uses the averaged background noise levels for all the microphones along a path [1]. It was assumed that the averaged background noise level will not vary much due to the absence of one microphone recording, i.e., for the position that may be different between the two repeated measurements in Type1. Despite its smaller sample size, the data for $L_{p,A,B}$ was analysed using the same method as the other ISO 3382-3 metrics, and only for Type1 repeats (see Section 2.5).

---

[4] For calculating STI, ISO 3382-3 recommends using IEC 60268-16 [6], which is the current (Edition4, dated 2011) STI standard [6], and genderless speech spectrum from ANSI 3.5 [7]. However, there is currently some ambiguity in ISO 3382-3 regarding STI calculation, since IEC 60268-16 [6] does not provide genderless octave-band weights and redundancy factors [6]. This was investigated in Cabrera et al. [8], and the results show that the choice of octave-band gender weighting used for STI calculations (from [6]) does not significantly affect the calculated $r_D$ values. This provides reassurance that the genderless speech spectrum specified in ISO 3382-3 is not compromised by using the male weights in STI calculation. Using male weights is implied but currently not specifically stated in ISO 3382-3, which perhaps should be clarified in future

[5] Although two offices were measured with three repeated measurement paths each, one path out of these was discarded for further analyses on Type2 repeats, in order to be consistent with the Type2 data from the rest of the five offices, where only two paths were possible. The discarded measurement paths are, nevertheless, included in Tables 1 and 2, since they are still valid as Type1 repeats.

**Table 1**
Relevant characteristics of 36 paths measured in 27 offices in 8 buildings labelled A-I, where offices with both Type1 and 2 repeated measurements (Section 2.3) are highlighted in bold. For complicated ceiling types, two heights are presented – to the closest substantial structures (support structures, exposed ductwork, etc.), and to the apex of the ceiling. For partition types: (N) represents little to no partition except computer screens, (I) represents a single partition separating adjacent workstations, (II) represents two partitions arranged in parallel or in a L, or a similar shape that encloses workstations from two sides, (III) represents three partitions that enclose the workstations from three sides, (IV) represents a single partition that does not extend all the way to the floor. An asterisk on a partition type denotes the type and height combination that was the most common for that office, amongst other combinations.

| Building (Office. Path) | Ceiling height and type | Floor Type | Length (m) | Width (m) | Workstations per 100 m² | Partition height (m), type |
|---|---|---|---|---|---|---|
| A (1.1) | 3.2; concrete | concrete | 28.2 | 25.5 | 16 | (N) |
| A (2.2) | 3.2; concrete | concrete | 27.0 | 26.0 | 16 | (N) |
| B (3.3) | 2.8; absorptive | carpeted | 22.0 | 8.0 | 4 | 1.1 (I, II*, III) |
| B (4.4) | 2.8; absorptive | carpeted | 26.7 | 7.5 | 4 | 1.3 (I, II*, III), 1.6 (I) |
| B (5.5) | 2.8; absorptive | carpeted | 21.8 | 5.8 | 4 | 1.3 (I, II, III) |
| C (6.6) | 3.7, 2.5; complicated: mostly vaulted, glass (70%), concrete (30%) | carpeted | 19.6 | 7.4 | 6 | 1.4 (I, II, III) |
| D (7.7) | 2.4, 7.6; vaulted: steel (80%), wood (10%), concrete (10%) | timber | 52.8 | 9.8 | 8 | (N) |
| D (8.8) | 2.7, 7.6; vaulted: steel (80%), wood (10%), concrete (10%) | timber | 52.8 | 9.8 | 8 | (N) |
| E (9.9) | 2.7; absorptive | carpeted | 18.4 | 9.0 | 12 | (N) |
| E (10.10) | 2.7; absorptive | carpeted | 25.0 | 9.4 | 11 | (N) |
| E (11.11) | 2.7; absorptive | carpeted | 17.3 | 11.7 | 10 | (N) |
| F (12.12) | 2.9; absorptive | carpeted | 15.4 | 10.0 | 9 | 1.2 (I) |
| F (13.13) | 3, 4.1; concrete (70%), absorptive (30%) | carpeted | 15.4 | 10.0 | 9 | 1.2 (I) |
| F (14.14) | 3; absorptive | carpeted | 12.3 | 4.3 | 6 | (N) |
| F (15.15) | 3, 4.1; absorptive (60%), concrete (40%) | carpeted | 10.6 | 5.7 | 9 | (N) |
| F (16.16) | 3.5; concrete (80%), absorptive (20%) | carpeted | 16.3 | 8.6 | 6 | 1.2 (I) |
| F (17.17) | 2.8; absorptive | carpeted | 17.4 | 6.6 | 8 | 1.2 (I) |
| **G (18.18-20)** | 2.7; absorptive | concrete | 11.5 | 7.0 | 18 | (N) |
| G (19.21) | 2.7; absorptive | concrete | 13.0 | 8.0 | 16 | (N) |
| G (20.22) | 2.7; absorptive | concrete | 17.0 | 8.6 | 15 | (N) |
| **G (21.23-25)** | 2.7; absorptive | concrete | 15.0 | 9.4 | 24 | (N) |
| **G (22.26-27)** | 2.7; absorptive | carpeted | 18.8 | 8.2 | 15 | (N) |
| **G (23.28-29)** | 2.7; absorptive | carpeted | 17.0 | 14.0 | 24 | 0.6 (IV) |
| **G (24.30-31)** | 2.7; absorptive | carpeted | 17.0 | 13.5 | 20 | 0.6 (IV) |
| H (25.32) | 2.7; absorptive | carpeted | 10.0 | 8.8 | 15 | 1.3 (I, II*) |
| **H (26.33-34)** | 2.7; absorptive | carpeted | 20.0 | 12.5 | 18 | 1.3 (I, II*) |
| **H (27.35-36)** | 2.7; absorptive | carpeted | 19.6 | 10.3 | 16 | 1.3 (I, II*) |

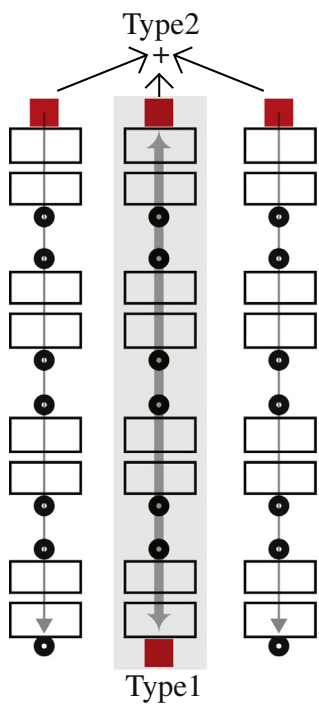

**Fig. 1.** Idealized Type1 and Type2 (which can include Type1 measurements) measurement paths, where the circles, squares, and rectangles represent the microphone, loudspeaker, and workstation locations, respectively.

### 2.4. Measurement conditions

All the measurements were supervised and conducted by the first author, with assistance from others. Per office, the repeatability conditions included measurements conducted by the same operator over a short time period, using the same method, and the same apparatus. Hence, given the constraints of the definition of an 'office' (see Sections 1 and 2.1), the guidelines in ISO 5725 1-6 [10], especially regarding repeatability conditions, were followed as closely as possible.

### 2.5. Statistical analyses

All the analyses were done within the software R [11], using the packages *tidyverse* (data management and plots) [12], *boot* [13], *robustbase* [14], and *robustlmm* [15]. Robust statistical methods were used throughout to reduce the influence of outliers, which included robust linear regression, and mixed-effects models; and using bootstrapping for calculating confidence intervals (CIs). Bootstrapping is a non-parametric resampling procedure that involves generating an empirical sampling distribution of the statistic of interest (e.g., mean), and using it for statistical estimation and inference, instead of making distributional assumptions [16,17]. In the following, wherever the use of bootstrapping is indicated, $10^5$ bootstrap resamples (each resample the size of the original data, and resampling with replacement) were generated, using the function *boot*, and the percentile CIs [18] calculated using the function *boot.ci* are reported; both functions are from the package *boot* [13]. Broadly speaking, analyses were performed on the difference between the largest and smallest observed values – the range – per metric, per Type1 and Type2 repeated measurement, which are referred to as ΔType1 and ΔType2, respectively (Section 2.5.1); and on the observed values of the metrics (Section 2.5.2).

### 2.5.1. Analysis using ΔType1 and ΔType2

The summary statistics included the mean and bootstrapped CIs (68% and 95%) of ΔType1 and ΔType2, where the mean values were calculated as,

$$\overline{\Delta \text{Type1}} = \frac{1}{36} \sum_{i=1}^{36} \Delta \text{Type1}_i \quad (1)$$

$$\overline{\Delta \text{Type2}} = \frac{1}{7} \sum_{j=1}^{7} \Delta \text{Type2}_j \quad (2)$$

where $i$ and $j$ are the path, and office indices, respectively (see Tables 1 and 2). Furthermore, for each metric, each $\Delta \text{Type1}_i$ and $\Delta \text{Type2}_j$ values were compared against the means $\left( \overline{\text{Type1}_i}, \overline{\text{Type2}_j} \right)$ of their respective underlying observed values; the latter (i.e., the means) being indicative of the magnitude of the metric for the $i$-th, and $j$-th index, respectively. These comparisons were made using the corresponding effect sizes for robust linear regression, i.e. $R^2$ values (referred to as $R^2_{\Delta \sim \text{mean}}$ in the following) and its 95% CI; and the significance (using 95% CIs) of the regression slopes, calculated using the *lmrob* function in the *robustbase* package [14]. Also, to determine the mutual relationship *between* the metrics for their respective ΔType1 and ΔType2 values, Kendall's $\tau$ correlation coefficients, along with bootstrapped 95% CIs were used.

### 2.5.2. Analysis using observed values of metrics

An important consideration for repeated measurements is *reliability*, which can be quantified using the intra-measurement correlation coefficient (ICC; generally known as intra-class correlation coefficient) [19] for both Type1 and Type2 repeated measurements, using the following equation:

$$\text{ICC} = \frac{(\sigma_B^2)}{(\sigma_B^2) + (\sigma_W^2)} \quad (3)$$

The ICC in Eq. (3) quantifies *reliability* as the relationship between the magnitude of the measurement error in the observed values per metric (i.e., within-measurement variance, $\sigma_W^2$), and the inherent variability in the 'error-free', or 'true' observed values of the repeated measurements per metric (i.e., between-measurement variance, $\sigma_B^2$). ICC has a value between 0 and 1, which is presented alongside its 95% CI in the current study. The idea is that if the reliability is high (i.e., high ICC), then the measurement errors are small in comparison to the true underlying variability between repeated measurements; in other words, the inherent, error-free variability in the repeated measurements per metric can be relatively well distinguished. Conversely, if the reliability is low (i.e., low ICC), measurements are error-prone, and the variability between repeated measurements could be predominantly due to the large underlying measurement errors, rather than due to actual variability between observed values. While there is no standard way to interpret ICC values, with interpretations varying between disciplines of research, the measurement reliability was classified according to common criteria [20] as excellent (ICC > 0.75), fair to good (ICC = 0.40–0.75) and poor (ICC ≤ 0.40). In Eq. (3), $\sigma_B^2$ and $\sigma_W^2$ were calculated by fitting a robust mixed-effects model per metric, using the *rlmer* function from the *robustlmm* package [15]. Each robust mixed-effects model, where the robustness corresponds to minimizing the contribution of outlying values [15], comprised of the fixed-effect of the respective metric (e.g., $r_D$), and the random-effects due to the repeated measurements in the offices.

To express the *repeatability*, several values could be provided, which are nevertheless dependent on the measurement error, i.e., within-measurement variability [19]. These values include $\sigma_W$, $\sqrt{k} \times \sigma_W$, where $k$ is the number of repeated measurements made per path/office ([10], part 6), and the repeatability coefficient (or, limit) $r$. The repeatability coefficient is calculated as:

$$r = f \times \sqrt{k} \times \sigma_W \quad (4)$$

The term $f$ in Eq. (4) depends on the probability level to be associated, and on the underlying probability distribution [10]. For an assumed normal distribution, and a probability level of 95%, the value of $f$ is 1.96. Hence, for $k$ = 2, the repeatability coefficient becomes $r = 2.8\sigma_W$, and the calculated value represents the upper limit within which 95% of the future differences between the Type1 repeated measurements are expected to occur. Similarly, for Type2 measurements, $k$ = 4, and $r = 3.6\sigma_W$ ([10], part 6).

## 3. Results and discussion

### 3.1. Acoustic summary of the repeated measurements in offices

The calculated acoustic metrics are presented in Table 2, where each row contains a path with a Type1 repeated measurement, with Type2 repeated measurements in seven offices (18, 21, 22, 23, 24, 26, 27; Table 2). To the best of the authors' knowledge, besides the studies by Virjonen et al. [21] and Haapakangas et al. [2], the current study presents the largest collection of paths/offices measured using the ISO 3382-3 method (Cabrera et al. [8] reported a subset of the current results), and is the first one that reports on more than one Type2 repeated measurements (Haapakangas et al. [2] reported results from repeated measurements in a single office described in Hongisto et al. [3]). The $R^2$ values for the linear regressions used to derive $r_D$ and $D_{2,S}$ [1] were fairly high: 0.94 ± 0.05, range: 0.75–1.00; and 0.95 ± 0.04, range: 0.85–1.00, respectively. As seen in Table 2, not many paths/offices measured are indicated to have 'good' acoustic conditions as per the calculated values of the metrics, including none for $r_D$, whereas plenty had 'poor' or 'insufficient' acoustic conditions [1].

### 3.2. Variability in ΔType1 and ΔType2

#### 3.2.1. ΔType1 results

All the $R^2_{\Delta \sim \text{mean}}$ values in Table 3, except for $r_D$, indicate some correlation between the $\Delta \text{Type1}_i$ values (i.e., difference per $i$-th path, per metric, in Table 2), and the corresponding $\overline{\Delta \text{Type1}_i}$ values (i.e., mean per $i$-th path, per metric in Table 2). However, the 95% CIs of their respective slopes crossed zero, i.e., are non-significant. In other words, overall, ΔType1 values did not vary significantly with the magnitude of the metric. With that in mind, the $\overline{\Delta \text{Type1}}$ values (Eq. (1)), which are based on measurements conducted with the loudspeaker on either end of the *same* path are fairly large, especially for both the $\text{SPL}_{\text{Speech}}$ metrics ($D_{2,S}$ and $L_{p,A,S,4\,m}$), and $L_{p,A,B}$ (Table 3; see also bootstrapped distributions in Fig. 2). However, the large $\overline{\Delta \text{Type1}}$ value for $L_{p,A,B}$ is perhaps less reliable due to the small sample size ($n$ = 13). Hence, the effect of $L_{p,A,B}$ on $r_D$ is not attempted here; moreover, $\overline{\Delta \text{Type1}}$ for $r_D$ is arguably not that large (∼1.2 m, Table 3). Fig. 2 shows a slight positive skew for the bootstrapped distributions for $\overline{\Delta r_D}$ and $\overline{\Delta L_{p,A,B}}$; however, the bias-corrected and accelerated (BCa) CIs that correct for skewness in bootstrapped distributions [22] were the same as the percentile CIs up to the first decimal. Hence, only the percentile CIs are reported here; see also [23] for evidence that supports the use of percentile over BCa CIs in some cases.

**Table 2**
Calculated ISO 3382-3 metrics and reverberation times for the all paths, where Type2 repeated measurements are highlighted in bold. The subscripts for the ISO 3382-3 metrics denote the index of the repeated measurement. The symbols '*' and '#' identify 'poor (or insufficient)' and 'good' values respectively based on the criteria in Annex A (informative) of ISO 3382-3 [1].

| Building (Office.Path) | $T_{30,mid}$ (s) | $L_{p,A,B}$ (dB) | $r_{D(1)}$ (m) | $r_{D(2)}$ (m) | $D_{2,S(1)}$ (dB) | $D_{2,S(2)}$ (dB) | $L_{p,A,S,4\ m(1)}$ (dB) | $L_{p,A,S,4\ m(2)}$ (dB) |
|---|---|---|---|---|---|---|---|---|
| A (1.1)    | 0.7 | 40.0 | 12.9* | 13.9* | 5.1  | 4.4*  | 53.4* | 54.0* |
| A (2.2)    | 0.7 | 37.0 | 16.5* | 17.6* | 4.8* | 2.6*  | 51.3* | 51.4* |
| B (3.3)    | 0.4 | 45.0 | 9.4   | 9.7   | 6.2  | 5.0   | 48.3* | 46.5  |
| B (4.4)    | 0.4 | 45.0 | 9.9   | 6.9   | 5.0  | 5.8   | 49.1* | 46.4# |
| B (5.5)    | 0.4 | 45.0 | 12.2* | 8.9   | 4.3* | 5.8   | 48.8* | 46.7# |
| C (6.6)    | 1.2 | 41.0 | 9.2   | 7.0   | 3.7* | 3.7*  | 51.7* | 50.9* |
| D (7.7)    | 0.7 | 41.0 | 8.3   | 9.2   | 4.8* | 5.4   | 47.5# | 48.0# |
| D (8.8)    | 0.6 | 38.0 | 11.9* | 11.0* | 5.1  | 5.5   | 46.0# | 46.2# |
| E (9.8)    | 0.5 | 46.0 | 7.0   | 7.7   | 4.0* | 4.8*  | 53.8* | 54.2* |
| E (10.10)  | 0.7 | 40.0 | 12.1* | 9.7   | 5.0  | 4.1*  | 54.4* | 51.6* |
| E (11.11)  | 0.6 | 43.0 | 10.3* | 11.6* | 4.2* | 4.0*  | 53.2* | 54.8* |
| F (12.12)  | 0.5 | 38.0 | 11.5* | 11.0* | 5.8  | 5.2   | 51.8* | 50.3* |
| F (13.13)  | 0.5 | 44.0 | 8.7   | 7.9   | 5.0  | 5.4   | 49.0* | 49.2  |
| F (14.14)  | 0.5 | 47.0 | 9.3   | 8.5   | 3.7* | 3.0*  | 53.3* | 53.0* |
| F (15.15)  | 0.4 | 54.0 | 8.7   | 9.4   | 3.6* | 2.7*  | 56.0* | 55.2* |
| F (16.16)  | 0.5 | 46.0 | 10.7* | 9.1   | 5.5  | 3.8*  | 52.4* | 49.0  |
| F (17.17)  | 0.4 | 46.0 | 9.0   | 9.0   | 6.2  | 7.0#  | 50.7* | 51.7* |
| **G (18.18)** | **0.4** | **42.0** | **8.4**  | **8.5**  | **4.6*** | **4.6*** | **52.0*** | **52.4*** |
| **G (18.19)** | **0.4** | **42.0** | **8.5**  | **7.6**  | **4.0*** | **4.5*** | **52.2*** | **51.8*** |
| **G (18.20)** | **0.4** | **42.0** | **8.7**  | **10.5*** | **4.7*** | **5.8** | **52.3*** | **54.3*** |
| G (19.21)  | 0.4 | 48.0 | 6.8   | 6.4   | 4.8* | 3.8*  | 52.7* | 52.8* |
| G (20.22)  | 0.4 | 45.0 | 8.0   | 8.0   | 6.2  | 5.6   | 55.1* | 53.3* |
| **G (21.23)** | **0.4** | **44.0** | **7.1**  | **6.9**  | **4.9*** | **5.0** | **53.7*** | **53.6*** |
| **G (21.24)** | **0.4** | **44.0** | **5.1**  | **7.3**  | **4.3*** | **4.1*** | **52.6*** | **53.9*** |
| **G (21.25)** | **0.4** | **44.0** | **7.6**  | **8.2**  | **4.0*** | **3.1*** | **52.4*** | **53.4*** |
| **G (22.26)** | **0.4** | **44.0** | **11.8*** | **10.2*** | **5.7** | **5.4** | **53.8*** | **53.4*** |
| **G (22.27)** | **0.4** | **44.0** | **9.6**  | **10.1*** | **4.8*** | **7.1#** | **51.4*** | **55.6*** |
| **G (23.28)** | **0.3** | **40.0** | **12.2*** | **11.5*** | **6.6** | **6.3** | **52.7*** | **52.3*** |
| **G (23.29)** | **0.3** | **40.0** | **13.1*** | **14.2*** | **5.3** | **5.8** | **51.1*** | **52.4*** |
| **G (24.30)** | **0.4** | **38.0** | **10.6*** | **10.3*** | **6.6** | **6.0** | **52.9*** | **52.8*** |
| **G (24.31)** | **0.4** | **38.0** | **11.4*** | **12.3*** | **6.4** | **5.4** | **52.9*** | **52.3*** |
| H (25.32)  | 0.4 | 45.0 | 7.7   | 8.7   | 5.4  | 6.5   | 50.5* | 52.4* |
| **H (26.33)** | **0.4** | **40.0** | **11.5*** | **12.0*** | **8.7#** | **8.3** | **52.8*** | **52.3*** |
| **H (26.34)** | **0.4** | **40.0** | **8.8**  | **9.7**  | **6.9** | **6.9** | **47.3#** | **47.9#** |
| **H (27.35)** | **0.3** | **35.0** | **13.7*** | **12.0*** | **8.4#** | **7.3#** | **52.4*** | **48.7** |
| **H (27.36)** | **0.3** | **35.0** | **8.9**  | **9.7**  | **7.0#** | **7.0#** | **54.1*** | **53.9*** |

**Table 3**
Results for analyses on ΔType1 and ΔType2 values, as described in Section 2.5.1.

| Type of repeated measurement | Metric | $\overline{\Delta Type1\ or\ 2}$, bootstrapped 95%, and 68% CIs | $R^2_{\Delta\sim mean}$, associated slope, and respective 95% bootstrapped CIs |
|---|---|---|---|
| Type1 (n = 36, except for $L_{p,A,B}$, where n = 13) | $r_D$ (m) | 1.17 (0.88,1.51), (1.02,1.33) | <$10^{-2}$ ($10^{-5}$,0.15), 0.15 (−0.35,0.83) |
| | $D_{2,S}$ (dB) | 0.78 (0.59,0.98), (0.68,0.88) | 0.11 ($10^{-2}$,0.35), −0.38 (−0.70,0.01) |
| | $L_{p,A,S,4\ m}$ (dB) | 1.16 (0.82,1.53), (0.98,1.34) | 0.15 ($10^{-2}$,0.40), −0.40 (−0.89,0.03) |
| | $L_{p,A,B}$ (dB) | 1.20 (0.72,1.73), (0.92,1.44) | 0.21 ($10^{-3}$,0.72), −2.13 (−3.97,1.05) |
| Type2 (n = 7) | $r_D$ (m) | 2.57 (1.79,3.42), (2.15,2.99) | 0.21 ($10^{-3}$,0.72), 0.83 (−0.12,2.69) |
| | $D_{2,S}$ (dB) | 1.36 (0.98,1.77), (1.16, 1.57) | 0.50 ($10^{-4}$,0.95), 1.60 (−0.95,3.48) |
| | $L_{p,A,S,4\ m}$ (dB) | 2.74 (1.33,4.27), (1.97, 3.49) | 0.28 ($10^{-4}$,0.71), −0.24 (−0.69,0.29) |

Fig. 3 presents another way to consider the ΔType1 values per metric, which are plotted with respect to the office floor area, and grouped according to the workstation density. Although there are no obvious global trends in Fig. 3, several paths had a large ΔType1 value for $L_{p,A,S,4\ m}$, and to an extent $D_{2,S}$, which implies that the direction of the measurement along a path is likely to be important in some cases, and needs to be judiciously considered during a measurement. Furthermore, in terms of mutual correlations *between* metrics for their respective $\Delta Type1_i$ values: $\Delta r_D$ was not significantly correlated to either $\Delta D_{2,S}$ ($\tau = 0.07; (-0.19, 0.30)$), or to $\Delta L_{p,A,S,4\ m}$ ($\tau = 0.2; (-0.02, 0.46)$); however, $\Delta D_{2,S}$ was significantly correlated to $\Delta L_{p,A,S,4\ m}$ ($\tau = 0.32; (0.03, 0.54)$). The latter, although not surprising given the underlying overlap in the calculation of the SPL$_{Speech}$ metrics, is nonetheless noteworthy, and can be expressed as the following robust linear model:

$$\Delta D_{2,S(\Delta Type1)} = 0.33 + 0.31 \times \Delta L_{p,A,S,4\ m(\Delta Type1)} \tag{5}$$

The $R^2$ = 0.46, 95% CI: (0.16,0.79) for Eq. (5), and the 95% CIs for the intercept and slope are (0.12,0.50) and (0.16,0.47), respectively.

Note that as per the definition of a path in Section 2.1, each path with large ΔType1 values (e.g., 16, 27, 35 in Fig. 3 for $L_{p,A,S,4\ m}$) was relatively uniform in relevant properties such as global absorptive treatments, ceiling heights, furniture arrangement (this is perhaps the weakest link here), etc. This was further confirmed by post-analysis examination of photographs and written notes by the operator for these measurements. Besides, for paths 27 and 35, other paths were measured within the same office, which were paths 26 and 36 respectively (see Table 2 and Fig. 4); here paths 26 and 36 had substantially smaller ΔType1 values for $L_{p,A,S,4\,m}$. Similarly, other large ΔType1 values for all the metrics in Fig. 3 were checked for any irregularities, and while the possibility of recording artefacts cannot be ruled out, there were no obvious errors in these measurements – future investigations are recommended to study such large ΔType1 in detail. Since the calculation of the SPL$_{Speech}$ based parameters does not involve using $L_{p,A,B}$ [1], its contribution to the noticed divergence can be ruled out. Note that $L_{p,A,B}$ is presumably also not a cause of divergence for $r_D$, because the same averaged spectrum is used for Type1 repeats; hence, cause of divergence is mostly some aspect(s) of room acoustics. Overall, it can be surmised that the SPL$_{Speech}$ based parameters seem more susceptible to a wide variation per Type1 repeats, and any averaging of these values must be done with caution.

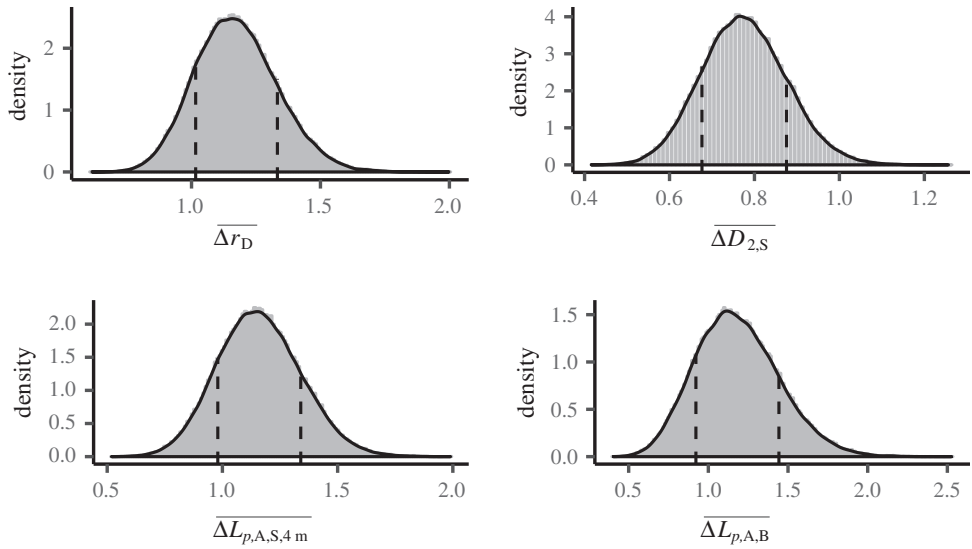

**Fig. 2.** Distribution of bootstrapped mean values for ΔType1 for the ISO 3382-3 metrics studied in the current paper (histogram binwidth = 0.01, density curve overlaid per metric). The bootstrapped 68% confidence interval (corresponding values are in Table 3) presented as vertical dashed lines per subplot.

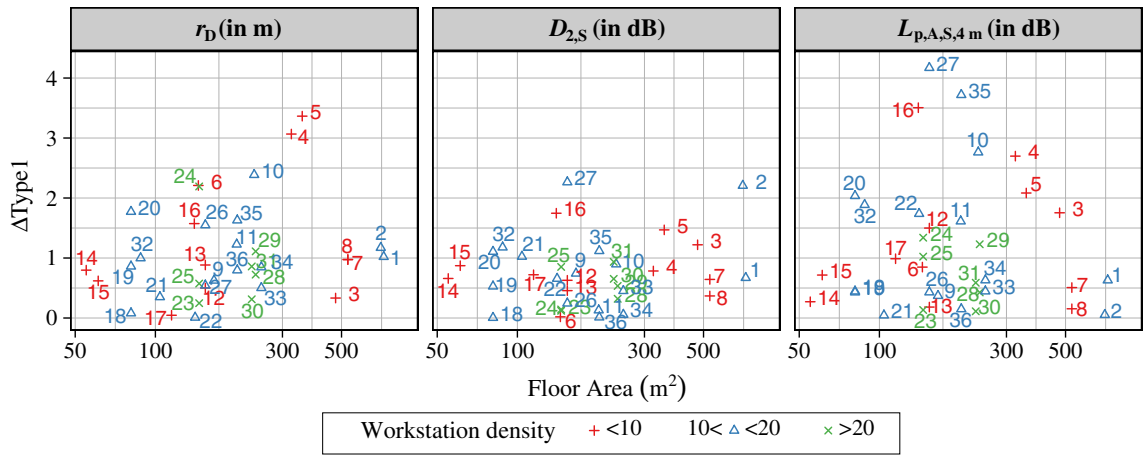

**Fig. 3.** ΔType1 for each path in Table 2, plotted against the office floor area, where the offices are grouped according to their workstation densities.

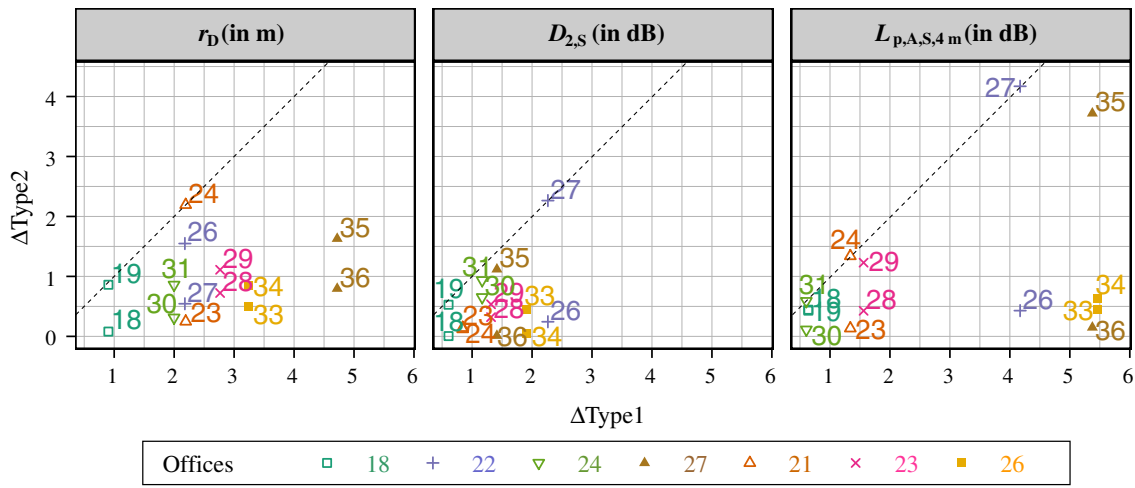

**Fig. 4.** The relationship between the ΔType1 and ΔType2 for the offices where both these types of measurements were performed (see Tables 1 and 2 for details about office and path numbers). The points plotted depict the path numbers, with the dashed line depicting the line with slope of 1 and intercept of 0.

### 3.2.2. ΔType2 results

Given the small sample size for ΔType2 ($n = 7$), the corresponding results in Table 3 are at best considered preliminary/exploratory at this stage, despite the use of bootstrapping (which still requires a fairly representative sample to begin with), and are discussed briefly alongside some individual ΔType2$_j$ measurements presented in Table 2.

The 95% CIs of the slopes crossing zero in Table 3 show that ΔType2 did not vary significantly with the magnitude of the metric, although the $R^2_{\Delta\sim mean}$ values are fairly high, especially for $D_{2,S}$. Overall for Type2 repeats, as seen in Table 3, $\overline{\Delta Type2}$ (Eq. (2)) are quite substantial for all the metrics, arguably to the extent of practically rendering most of the underlying measurements unusable for characterizing offices without further investigation. Although the mutual correlations *between* metrics for their respective ΔType2$_j$ values are provided here, these correlations must not be used before more exhaustive investigations are conducted. With that in mind, $r_D$ was not significantly correlated to either $D_{2,S}$ ($\tau = 0.18; (-0.26, 0.58)$), or to $L_{p,A,S,4\,m}$ ($\tau = 0.37; (-0.02, 0.68)$); and, $D_{2,S}$ was not significantly correlated to $L_{p,A,S,4\,m}$ ($\tau = 0.37; (-0.01, 0.64)$).

More detailed observations are possible based on Fig. 4, where the $\overline{\Delta Type2}$ values are plotted in relation to the ΔType1 values in offices where both these measurements were performed. Clearly, except for perhaps offices 18 and 24, the rest of the offices in Fig. 4 were particularly variable in some aspect(s), as noticed by the large ΔType1 values; in some cases, values for ΔType2 ≤ΔType1, though not by much. On the first glance, this seems to be at odds with the values in Table 3 (third column); however, note that individual offices are being considered here, whereas the sample size for the statistics in Table 3 is substantially larger for the Type1 repeats, compared to Type2 repeats. Continuing with individual comparisons based on Fig. 4, in three offices (24, 27, and 31), the values for ΔType1 = ΔType2 for at least one of the metrics plotted, which can be seen by the intersection of ΔType1/2 values with the line of slope 1, intercept 0. Hence, for these three offices, one of the ΔType1 values out of two that contributed towards the ΔType2 value in the office had a range larger than the other; the latter being a different measurement path in the same office. This suggests that the current method of using only two paths for characterising Type2 repeats is potentially insufficient – more than two, and perhaps multiple paths are recommended, wherever possible, for Type2 repeats. Otherwise, caution must be exercised in characterising offices using Type2 repeats. Furthermore, for an arbitrary split at a value of 1 for all the metrics, out of 42 values in Fig. 4, 10 have ΔType2 ≥1 (~24%; more-or-less inconsistent with the CIs for ΔType1 in Table 3), and 10 (~24%) have ΔType1 & ΔType2 ≤1. The former perhaps represents large values, especially for SPL$_{Speech}$ metrics for ΔType2, with the former and latter in combination highlighting the fact that quite a few of the ΔType1 values in offices in Fig. 4 were fairly large (see also Section 3.2.1). $L_{p,A,S,4\,m}$ was especially large in two cases – path 35 and 27, which were also discussed in Section 3.2.1; also see Section 3.3 where the low *reliability* of $L_{p,A,S,4\,m}$ Type2 repeats is discussed. In summary, more studies are needed with larger sample size of offices with Type2 measurements to determine the relevance of the findings in the current study as discussed above, and presented in Fig. 4.

Moreover, the high values of ΔType1 and ΔType2, in general, raise some questions that may be relevant for future enhancements and/or clarifications in the ISO 3382-3 methodology. Given the plethora of variations in office topologies, both geometrically and in terms of furniture arrangement including absorptive elements, and the associated room acoustics variations that can be both large and subtle, ISO 3382-3 perhaps needs to be very specific in what would constitute a repeated measurement that is *not* a Type1 repeat. Furthermore, there might need to be a threshold for tolerance of ΔType2 and perhaps also ΔType1, beyond which measurements cannot be used to characterize an office. Determining just-noticeable differences, especially for $r_D$, might also be beneficial. Addressing these, and similar questions and improvements will, of course, require large-scale, focussed studies in the future.

As an aside, a comment can be made about the ISO 3382-3 requirement of keeping the loudspeaker and measurement positions at least 2 m away from large reflective surfaces [1]. This requirement, which is of course important from a wave theoretical perspective, limits the use of ISO 3382-3 method to mostly medium- to large-sized offices. The current finding of large ΔType2 (and the underlying ΔType1) values provides a practice-based evidence in favour of '2 m rule', which would otherwise be a confounding factor in critically assessing results related to the ISO 3382-3 method in the current, and in future studies.

### 3.3. Variability in the observed Type1 and Type2 values

Based on the discussion in Section 3.2, where it was shown that the findings about Type2 repeated measurements are at best exploratory, and require future studies, the focus in this section will be on Type1 repeated measurements that have larger sample size.

### 3.3.1. Reliability

ICC values above 0.8 are generally considered between very-good to excellent [20] in terms of reliability, which is applicable to all the metrics and both Types 1 and 2, as seen in Table 4, except for $L_{p,A,S,4\,m}$ for Type2 repeats, which had low ICC, and hence, poor reliability (also non-significant). The latter provides another perspective on the large ΔType2 values, as discussed in Section 3.2.2, and reinforces the need to further investigate, and define the scope of Type2 repeated measurements. Also note that the $L_{p,A,B}$ ICC results are based on thirteen samples, which can be considered a small sample from a statistical perspective. Overall, the high reliability of the ISO 3382-3 metrics implies that the current findings are primarily addressing the true variability between the underlying values, with acceptable measurement errors – providing support for the critical assessment of the ISO 3382-3 measurement method in the current, and similar studies in the future.

### 3.3.2. Repeatability

The last two columns of Table 4 provide the relevant repeatability metrics. In general, the standard deviations reported here do not seem unreasonably large, especially for $r_D$. The only previous data in this regard was provided by Haapakangas et al. [2], where the variability[6], expressed as the 68% confidence interval in the ISO 3382-3 metrics, was estimated as ±1 dB for $D_{2,S}$, ±1.5 dB for $L_{p,A,S,4,m}$, ±1.5 m for $r_D$, and ±1 dB $L_{p,A,B}$. These variability estimates were based on unpublished data about repeated measurements within a single open-plan office (described in Hongisto et al. [3]), which is similar in principle to the current Type2 repeats, although there is likely to be a difference in the respective methods (i.e., current and in Hongisto et. al. [3]), including the number of repeats. Besides including two types of repeated measurements, the current research includes measurements within several offices, which allows providing variability estimates that are generalizable to a wider range of offices. Nevertheless, compared to Haapakangas et al. [2], the standard deviations in the current study ($\boldsymbol{\sigma_W}$ in Table 4) were smaller for $D_{2,S}$ (for both Type1 and Type2 repeats), for $L_{p,A,S,4\,m}$ (for Type1 repeats; Type2 unreliable), and for $r_D$ (for both Type1 and Type2 repeats); and was almost the same for $L_{p,A,B}$ (for Type1 repeats).

**Table 4**
Some reliability (intra-measurement correlation coefficient: **ICC**) and repeatability metrics (within-measurement standard deviation: $\sigma_W$; and repeatability limit/coefficient: $r$) for both Type1 and Type2 repeated measurements.

| Type of repeated measurement | Metric | ICC (95% CI) | $\sigma_W$ | $r$ |
|---|---|---|---|---|
| Type1 ($n$ = 36, except for $L_{p,A,B}$, where $n$ = 13) | $r_D$ (m) | 0.87 (0.77,0.93) | 0.90 | 2.5 |
| | $D_{2,S}$ (dB) | 0.85 (0.73,0.92) | 0.61 | 1.7 |
| | $L_{p,A,S,4\,m}$ (dB) | 0.82 (0.68,0.90) | 1.04 | 2.9 |
| | $L_{p,A,B}$ (dB) | 0.97 (0.91, 0.99) | 1.01 | 3.0 |
| Type2 ($n$ = 7) | $r_D$ (m) | 0.80 (0.55,0.95) | 1.15 | 4.5 |
| | $D_{2,S}$ (dB) | 0.82 (0.56,0.96) | 0.44 | 1.7 |
| | $L_{p,A,S,4\,m}$ (dB) | 0.21 (−0.1, 0.71) | 1.62 | 6.3 |

The repeatability coefficient/limit ($r$) per metric provides a more exhaustive, and moreover, standardized account of the variability within repeated measurements. As described in Section 2.5.2, the $r$ value indicates the upper limit *not likely* to be exceeded in 95% of future instances, where the difference between observed values of repeated measurements are studied, either on the same path for Type1 repeats, or the same office for Type2 repeats. For instance, for 95% of the instances of Type1 repeats – for measurements by the same observer using the current method, on the same path measured with loudspeakers on either end – the ΔType1 for $L_{p,A,S,4\,m}$ is unlikely to be more than ∼ 3 dB ($r$ value in Table 4). Similar assessments can be made for the other metrics for Type1 repeats, whereas the $r$ values for Type2 repeats are fairly large, and are not recommended for any use other than informing future research in more offices, and with, moreover, a clearly-defined (perhaps standardized in ISO 3382-3) nature of Type2 repeated measurements.

Overall, the practical implications of the standard deviations, and repeatability limit/coefficients reported here, and whether an appraisal from a theoretical/methodological perspective is required, are open questions.

## 4. Conclusions

The current findings represent a substantial step towards quantifying the variability of the ISO 3382-3 metrics in open-plan offices under repeatability conditions. Specifically, for the two types of repeated measurements considered here, new results are provided for the *repeatability* and *reliability* of the ISO 3382-3 metrics, along with relevant summary statistics for the range of the values of the ISO 3382-3 metrics. The summary statistics, and repeatability values reported here may be considered reasonable for the most part, especially for $r_D$; the variability of the SPL$_{Speech}$ metrics may need to be considered further from a practical perspective. In this regard, the excellent reliability values reported here imply that future investigations with repeated measurements are more likely to be studying the 'true' variability between ISO 3382-3 metrics in offices, with acceptable measurement error.

For the Type1 repeats, the large and moreover diverse sample of offices underlines the usefulness of the current results in terms of characterising a wide range of room acoustic environments in offices. However, for the SPL$_{Speech}$ metrics ($D_{2,S}$ and $L_{p,A,S,4\,m}$) in particular, and $r_D$ to lesser extent, large ΔType1 in some cases demonstrated variability being dependent on the direction of the source-receiver configuration in the measurement path. This provides some context, and more importantly, reinforcement for the ISO 3382-3 requirement of at least measuring a path from two opposite directions (i.e., Type1 repeat), if it is the only path being measured. Furthermore, based on the current results, it is highly recommended that *all* paths be measured from both ends. For exceptionally large ΔType1 values that are otherwise ISO 3382-3 compliant, it would be apt to flag such cases while reporting/using the results. ISO 3382-3 may benefit from amendments to its reporting format (Fig. 3/Table 2 in [1]) to highlight such cases, and perhaps more crucially, include instructions/comments on the relevance and potential remedies to large ΔType1 from a practitioner's perspective.

The practical applicability of Type2 results is currently limited due to small sample size in this paper. Nevertheless, some of the insights into Type2 repeats from individual assessments of offices in sections 3.2.2 and 3.3 may be helpful in informing future studies, and consequently, potential updates to the ISO 3382-3 methodology regarding latent issues in Type2, e.g., clarity in its specification, etc. More focussed studies may also be beneficial where the influence of specific room acoustic issues, including furniture density, absorption arrangement, etc. on variability in metrics are studied (ideally) in a factorial design.

Ultimately, more work is needed, perhaps from inter-laboratory, and international coordination, to further strengthen ISO 3382-3 in relation to measurement *repeatability* (from current and future findings based on [10]), other aspects of variability in repeated measurements including *reproducibility* (based on [10]), and perhaps even *uncertainty* at various stages in the methodology (based on [24], similar to [25]).

## Acknowledgments

This study was funded through the Australian Research Council's Discovery Projects scheme (project DP160103978). The authors thank Daniel Protheroe (Marshall Day Acoustics) for assistance with some aspects of this paper.

---

[6] Note that Haapakangas et al. [2] refer to the 'uncertainty' of the metrics, whereas the term 'variability' is preferred here. This is to avoid confusion with the more elaborate exercise in determining the 'uncertainty' in measurement results [24]. Using the term variability also signifies the ongoing nature of the contribution of this paper towards the uncertainty issue, which may ultimately lead to a more robust framework regarding the uncertainty in the ISO 3382-3 measurements, and metrics.

146	M. Yadav et al. / Applied Acoustics 150 (2019) 138–146## References

[1] ISO 3382-3 Acoustics – measurement of room acoustic parameters – Part 3: Open plan offices. International Organization for Standardization, Geneva, Switzerland; 2012.
[2] Haapakangas A, Hongisto V, Eerola M, Kuusisto T. Distraction distance and perceived disturbance by noise—an analysis of 21 open-plan offices. J Acoust Soc Am 2017;141:127–36. https://doi.org/10.1121/1.4973690.
[3] Hongisto V, Haapakangas A, Varjo J, Helenius R, Koskela H. Refurbishment of an open-plan office–environmental and job satisfaction. J Environ Psychol 2016;45:176–91. https://doi.org/10.1016/j.jenvp.2015.12.004.
[4] Cabrera D, Jimenez D, Martens WL. Audio and Acoustical Response Analysis Environment (AARAE): a tool to support education and research in acoustics. Proc. Internoise, Melbourne, Australia: 2014.
[5] ISO 3745 Acoustics – Determination of sound power levels and sound energy levels of noise sources using sound pressure – Precision methods for anechoic rooms and hemi-anechoic rooms. International Organization for Standardization, Geneva, Switzerland; 2012.
[6] IEC 60268-16 Ed.4, Sound system equipment - Part 16: Objective rating of speech intelligibility by speech transmission index. International Electrotechnical Commission, Geneva, Switzerland; 2011.
[7] ANSI 3.5-1997 (R2012). American National Standard – Methods for Calculation of the Speech Intelligibility Index. Acoustical Society of America; 2012.
[8] Cabrera D, Yadav M, Protheroe D. Critical methodological assessment of the distraction distance used for evaluating room acoustic quality of open-plan offices. Appl Acoust 2018;140:132–42. https://doi.org/10.1016/j.apacoust.2018.05.016.
[9] Cabrera D, Xun J, Guski M. Calculating reverberation time from impulse responses: a comparison of software implementations. Acoust Aust 2016;44:369–78. https://doi.org/10.1007/s40857-016-0055-6.
[10] ISO 5725 1-6 Accuracy (trueness and precision) of measurement methods and results. International Organization for Standardization, Geneva, Switzerland; 1994.
[11] R Core Team. R: A Language and Environment for Statistical Computing. Vienna, Austria. URL https://www.R-Project.Org/: R Foundation for Statistical Computing; 2018.
[12] Wickham H. Tidyverse: Easily install and load 'tidyverse' packages. R package version 1.2.1. https://CRAN.R-project.org/package=tidyverse 2017.
[13] Canty A, Ripley BD. boot: Bootstrap R (S-Plus) Functions. R package version 1.3-20. 2017.
[14] Maechler M, Rousseeuw P, Croux C, Todorov V, Ruckstuhl A, Salibian-Barrera M, et al. robustbase: Basic Robust Statistics. R package version 0.93-3. URL http://CRAN.R-project.org/package=robustbase. 2018.
[15] Koller M. robustlmm: an r package for robust estimation of linear mixed-effects models. J Stat Softw 2016;75:1–24. https://doi.org/10.18637/jss.v075.i06.
[16] Efron B, Tibshirani R. Bootstrap methods for standard errors, confidence intervals, and other measures of statistical accuracy. Stat Sci 1986;1:54–75.
[17] Efron B, Tibshirani RJ. An Introduction to the Bootstrap. CRC Press; 1994.
[18] Efron B. Nonparametric estimates of standard error: the jackknife, the bootstrap and other methods. Biometrika 1981;68:589–99. https://doi.org/10.1093/biomet/68.3.589.
[19] Bartlett JW, Frost C. Reliability, repeatability and reproducibility: analysis of measurement errors in continuous variables. Ultrasound Obstet Gynecol 2008;31:466–75. https://doi.org/10.1002/uog.5256.
[20] Fleiss JL. Design and Analysis of Clinical Experiments, 73. John Wiley & Sons; 2011.
[21] Virjonen P, Keränen J, Hongisto V. Determination of acoustical conditions in open-plan offices: proposal for new measurement method and target values. Acta Acust United Acust 2009;95:279–90. https://doi.org/10.3813/AAA.918150.
[22] Efron B. Better bootstrap confidence intervals. J Am Stat Assoc 1987;82:171–85. https://doi.org/10.1080/01621459.1987.10478410.
[23] Biesanz JC, Falk CF, Savalei V. Assessing mediational models: testing and interval estimation for indirect effects. Multivar Behav Res 2010;45:661–701. https://doi.org/10.1080/00273171.2010.498292.
[24] ISO, IEC, OIML, BIPM. Guide to the Expression of Uncertainty in Measurement. International Organization for Standardization, Geneva, Switzerland; 1995.
[25] Modeling Dietrich P. Measurement uncertainty in room acoustics. Acta Polytech 2007;47. https://doi.org/10.14311/972.